\documentclass[aps,twocolumn, floatfix,superscriptaddress]{revtex4-1}
\usepackage{graphicx}
\usepackage{hyperref}
\usepackage{amsmath}
\usepackage{amsfonts}
\usepackage{mathtools,bm}
\usepackage[amssymb]{SIunits}
\usepackage{color}
\usepackage{mathrsfs}
\usepackage{animate}

\begin{document}
\flushbottom
\title{Self-accelerating beam dynamics in the space fractional Schr\"{o}dinger equation}

\author{David Colas}
\email{d.colas@uq.edu.au}
\affiliation{ARC Centre of Excellence in Future Low-Energy Electronics Technologies, School of Mathematics and Physics, University of Queensland, St Lucia, Queensland 4072, Australia}

\begin{abstract}
Self-accelerating beams are fascinating solutions of the Schr\"{o}dinger equation. Thanks to their particular phase engineering, they can accelerate without the need of external potentials or applied forces. Finite-energy approximations of these beams have led to many applications, spanning from particle manipulation to robust  \textit{in vivo} imaging. The most studied and emblematic beam, the Airy beam, has been recently investigated in the context of the fractional Schr\"{o}dinger equation. It was notably found that the packet acceleration would decrease with the reduction of the fractional order. Here, I study the case of a general $n\textsuperscript{th}$-order self-accelerating caustic beam in the fractional Schr\"{o}dinger equation. Using a Madelung decomposition combined with the wavelet transform, I derive the analytical expression of the beam's acceleration. I show that the non-accelerating limit is reached for infinite phase order or when the fractional order is reduced to 1. This work provides a quantitative description of self-accelerating caustic beams' properties. 
\end{abstract}

 \date{\today} \maketitle

\section{Introduction}
The Schr\"{o}dinger equation has been at the heart of quantum mechanics for nearly a century. However, it is only two decades ago that this fundamental equation of physics has been extended to fractional calculus~\cite{herrmann_book14a}, thanks to Laskin~\cite{laskin00a,laskin00b,laskin02a}. He generalized Feynman's path integral formulation of quantum mechanics to L\'{e}vy flights, \textit{i.e.} beyond Brownian motion or Wiener stochastic processes that are based on usual Gaussian statistics. This led to establish the basis of fractional quantum mechanics and to derive the space fractional Schr\"{o}dinger equation (SFSE). 
Fractional derivatives in partial differential equations have a long history in accurately modeling a wide range of physical systems, where their integer-order counterparts failed~\cite{kilbas_book06a}. This includes systems experiencing anomalous diffusion, like fluids in heterogeneous porous media with long range spatial correlation decaying as a power law~\cite{metzler00a}. It also concerns several linear or nonlinear viscoelastic and wave propagation problems (acoustic, water...) in lossy media, or even industrial controllers for vehicule suspensions~\cite{sabatier_book07a}. Following Laskin's work, various schemes for a physical implementation of the SFSE have been discussed~\cite{stickler13a,longhi15a,zhangd17a}.

Amongst the many solutions of the Schr\"{o}dinger equation, the Airy beams discovered in 1979 by Berry and Balazs~\cite{berry79a} share a particular place. These intriguing wave packet solutions notably possess the apparent property of accelerating without external potentials or applied forces.  They were long considered to be a mathematical curiosity until their physical realisation, using approximations of Airy beams with a finite energy~\cite{siviloglou07a,siviloglou07b,ellenbogen09a}. They were subsequently observed in platforms other than optically-based ones, using \textit{e.g.} electron beams~\cite{voloch13a} or surface plasmon polariton~\cite{zhangP11a}, and used in applications such as optical particle manipulation~\cite{baumgartl08a}, light sheet microscopy~\cite{vettenburg14a} and many others~\cite{polynkin09a,abdollahpour10a,gu10a,nagar19a}. 

Fundamentally, accelerating beams can be understood in the broader context of the catastrophe theory, introduced by Thom~\cite{thom_book72a,berry80a}. Their caustic properties arise from diffraction integrals which define the different sets of catastrophe in Thom's theory. Aside from the widely studied Airy beam that connects from the fold catastrophe, only little attention was brought to caustic beams of higher order, such as Pearcey and Swallowtail beams~\cite{ring12a,zannotti17a,zangf19a}. These beams share similar accelerating properties but with a different acceleration parameter. Recently, Airy beams physics has been investigated in the context of the SFSE~\cite{huangx17a,huangx17b,zhangl19a}. For a free particle, it has been qualitatively shown that the acceleration of the Airy packet decreases as the fractional index is reduced. These last developments invite for a more general and quantitative description of self-accelerating beams in the SFSE.  

In this paper, I study the properties of general one-dimensional self-accelerating beams in the context of the SFSE. The wave packet acceleration along with the peaks' trajectory is obtained analytically using a simple method based on the phase decomposition of the wave function in momentum space~\cite{colas20a}. This method, combined with the use of the wavelet transform (WT), allows to understand self-accelerating beams as the result of a self-interference of the wave function, and to monitor the trajectory of their individual modes. This effect was notably found, following a similar approach, in exciton-polaritons~\cite{colas16a}, atomic condensates~\cite{colas18a} or to be at the origin of the formation of nonlinear X-waves~\cite{colas19a}.

The paper is organised as follows. In Section II, I introduce the formalism of the SFSE and I derive a general wave packet solution. In Section III, I introduce self-accelerating beams as derived from catastrophe theory. In Section IV, I present a detailed analysis of the Airy beam as solution of the SFSE. In Section V, I extend these results to the case of a general self-accelerating beam within the SFSE, deriving the general wave packet acceleration. Finally, Section VI concludes the paper. 
\section{Wave packet dynamics in the fractional Schr\"{o}dinger equation}
\label{sec:2}
Let's consider the one-dimensional time-dependent space fractional Schr\"{o}dinger equation as introduced by Laskin~\cite{laskin00a,laskin00b,laskin02a}
\begin{equation}
i\hbar \partial_t \psi(x,t) = - D_\alpha[\hbar\nabla]^\alpha \psi(x,t)\,,
\label{eq:SFSE_x}
\end{equation}
where $[\hbar\nabla]^\alpha$ is the so-called quantum Riesz derivative of fractional order $\alpha$, or L\'{e}vy index, satisfying $1<\alpha\leq 2$. The prefactor $D_\alpha$ is the quantum diffusion constant with dimension $[\mathrm{energy}]^{1-\alpha}[\mathrm{lenght}]^\alpha [\mathrm{time}]^{-\alpha}$. With $\alpha=2$ and $D_\alpha=1/2m$, $m$ being the mass of the particle, one recovers the conventional one-dimensional Schr\"{o}dinger equation. Using the definition of the Riesz derivative~\cite{bayin16a}, the SFSE can be more conveniently written in momentum space as
\begin{equation}
i\partial_t \psi(k,t)=D_\alpha |k|^\alpha \psi(k,t)\,,
\label{eq:SFSE_k}
\end{equation}
with $\hbar=1$. In Eq.~(\ref{eq:SFSE_x}), the fractional derivative in real space is intrinsically a non-local operator~\cite{bayin16a}. However, regarded from a solid-state physics point of view, it simply corresponds to a modification of the dispersion relation in momentum space, here defined as $E^{(\alpha)}(k)=D_\alpha |k|^\alpha$~\cite{note2}. It spans from a parabola $(\alpha=2)$ to a symmetric linear dispersion $(\alpha\rightarrow 1)$, much as the one experienced by conduction electrons on a Dirac cone~\cite{castronetoa09} or by Bogoliubov excitations in an interacting Bose-Einstein condensate~\cite{utsunomiya08a}.

For a given initial condition $\psi(k,0)$, the solution of  Eq.~(\ref{eq:SFSE_k}) is obtained by simple integration:
\begin{equation}
\psi(k,t)=\exp(-i E^{(\alpha)}(k) t )\psi(k,0)\,.
\end{equation}
In his seminal article~\cite{laskin00b}, Laskin solved Eq.~(\ref{eq:SFSE_k}) for a symmetric L\'{e}vy stable distribution, or ``L\'{e}vy packet'', that can be expressed analytically through its characteristic function, \textit{i.e.} its Fourier transform, as
\begin{equation}
\hat{L}_\mu(k) = \exp(-a_\mu|k|^\mu)\,,
\label{Levy_k}
\end{equation}
where $a_\mu$ is a constant scaling the amplitude of the distribution's (\textit{Pareto}) tails. Analytical expressions in position space for Eq.~(\ref{Levy_k}) only exist for the case $\mu=2$ where ${L}_2(x)$ is a Gaussian distribution, and for $\mu=1$ where ${L}_1(x)$ is a Lorenztian distribution. With such an initial condition, the solution of Eq.~(\ref{eq:SFSE_k}) for a L\'{e}vy packet reads
\begin{equation}
\psi(k,t)=\exp\left(-i D_\alpha|k|^\alpha t -a_\mu|k|^\mu\right)\,.
\end{equation}
\begin{figure}[t!]
  \includegraphics[width=\linewidth]{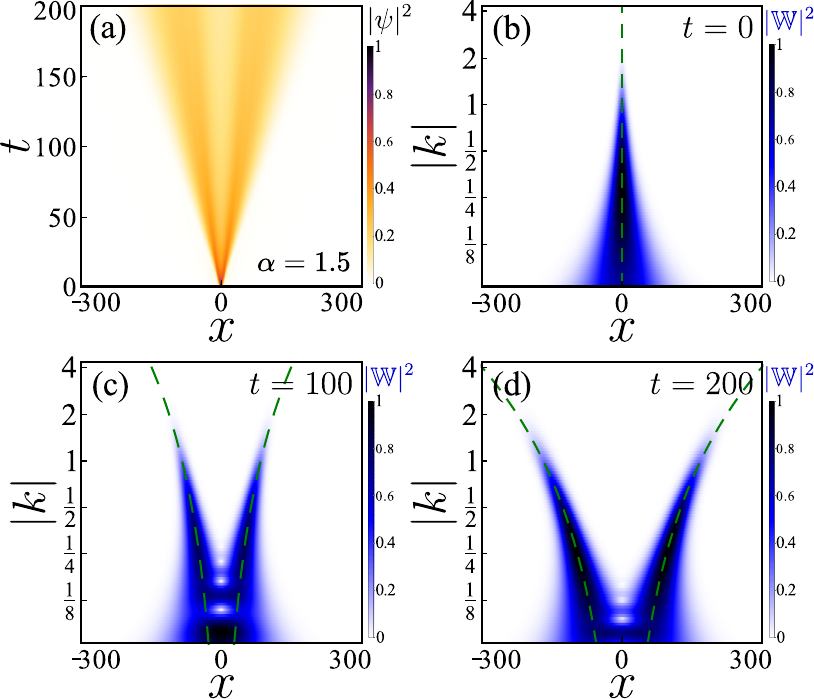}
  \caption{ Diffusion of a Gaussian wave packet in the SFSE of order $\alpha=1.5$, and with initial width $\sigma_k=1/1.5$. (a) Wave function density $|\psi(x,t)|^2$. (b)-(d) Wavelet energy density $|\mathbb{W}(x,k)|^2$ computed at selected times. The dashed green lines stand for the mode displacement $d^{(\alpha)}(k,t)$ computed from Eq.~(\ref{eq:dkt}) and derived from the dispersion relation only. Supplemental Movie S1 provides an animation of the Gaussian wave packet dynamics with its WT~\cite{noteSM}.
  }
  \label{fig:2}
\end{figure}

As I discuss in the following, this choice of the initial condition does not influence the mode propagation, which here only depends on the properties of the dispersion, but only the mode population through the spread of the wave function in momentum space.

I first consider the simple case of a Gaussian wave packet, \textit{i.e.} with $\mu=2$ in Eq.~(\ref{Levy_k}), as initial condition: $\psi(k,0)=\exp(-k^2/2\sigma_k^2)$. The solution of Eq.~(\ref{eq:SFSE_k}) is now
\begin{equation}
\psi(k,t)=\exp\left(-i D_\alpha|k|^\alpha t\right)\exp\left(-\frac{k^2}{2\sigma_k^2}\right)\,.
\label{eq:solGaussian}
\end{equation}
To analyse and understand the individual mode properties of the wave packet solution given in Eq.~(\ref{eq:solGaussian}), I employ the method developed in Ref.~\cite{colas20a}. It consists in studying the phase properties of the wave function $\psi(k,t)$, since only a phase dynamics takes place in momentum space. The wave function is decomposed as $\psi(k,t)= \sqrt{N(k)} \exp(-i \phi(k,t))$, with an amplitude $\sqrt{N}$ and a phase term $\phi$. This is analogous to a Madelung transformation in position space, notably performed in hydrodynamics studies~\cite{sonin_book16a} and where the gradient of the (real space) phase corresponds to the fluid velocity. In the present case, the initial condition (either Gaussian or L\'{e}vy) does not have a complex phase, so the whole phase dynamics arises as a consequence of the dispersion relation
\begin{equation}
\label{eq:phasek}
\phi(k,t)=D_\alpha|k|^\alpha t\,.
\end{equation}
The gradient of this phase, with respect to $k$, gives the wave packet's mode displacement~\cite{colas20a}
\begin{equation}
d^{(\alpha)}(k,t)=\alpha D_\alpha t |k|^{\alpha-1} \textrm{sgn}(k)\,.
\label{eq:dkt}
\end{equation}
This quantity has a simple interpretation, that a particle in a given mode $k$ will propagate at a distance $d^{(\alpha)}$ after a time $t$. This has been equivalently introduced by Laskin as the expectation value of the space position for a single mode $k_0$ of the dispersion relation~\cite{laskin00b}:
\begin{equation}
\langle x \rangle=\partial_k E^{(\alpha)}(k)\Bigr|_{\substack{k=k_0}}t=v_0 t\,,
\end{equation}
using the $k$-dependent group velocity dispersion
\begin{equation}
v(k)=\partial_k E^{(\alpha)}(k) = \alpha D_\alpha |k|^{\alpha-1} \textrm{sgn}(k)\,.
\end{equation}
Like for the phase in Eq.~(\ref{eq:phasek}), the mode displacement in Eq.~(\ref{eq:dkt}) is independent of the initial condition. The choice of $\psi(k,0)$ only governs the population of the modes, but not their propagation. A Gaussian distribution has a well-defined variance $\sigma_k^2$ so the populated modes are then well-located in a certain range of momentum, whereas a more general L\'{e}vy stable distribution has no defined variance (for $\mu <2$) due to the presence of \textit{Pareto} tails. The dynamics of $|\psi(x,t)|^2$ for a Gaussian wave packet in the SFSE of order $\alpha=1.5$ is shown in Fig.~\ref{fig:2}(a). Unlike the conventional case of $\alpha=2$, the packet splits into two entities at long times, which is a direct consequence of the modified dispersion.

To better visualise and understand the effect of the mode displacement $d^{(\alpha)}(k,t)$, I use the wavelet transform (WT) which permits an alternative representation of the wave function in both position ($x$) and momentum ($k$). The WT is defined as~\cite{debnath_book15a}
\begin{equation}
\mathbb{W}(x,k)=(1/\sqrt{|k|})\int_{-\infty}^{+\infty}\psi(x')\mathcal{G}^\ast [(x'-x)/k]\mathrm{d}x'\,,
\end{equation}
where I use Gabor wavelet family
\begin{equation}
\mathcal{G}(x)=\sqrt[4]{\pi}\exp(i w_\mathcal{G} x)\exp(-x^2/2)\,,
\end{equation}
which are Gaussian functions with an internal frequency $w_\mathcal{G}$. The choice of the Gabor wavelet is arbitrary but here natural since Gaussians are simple wave packet solutions of the SE. One could however use other wavelet families (Morlet, Mexican hat etc.) to obtain equivalent results. I apply the WT to the Gaussian packet previously calculated and show its corresponding wavelet energy densities $|\mathbb{W}(x,k)|^2$ at different times of the evolution in Fig.~\ref{fig:2}(b)-(d). The density is well-fitted by the mode displacement $d^{(\alpha)}(k,t)$, here shown as dashed green lines. This representation permits an efficient way to visualize, at a given instant of time, both the population of the modes and the distance they have propagated. 

As I will show in Section~\ref{sec:4}, the mode displacement can be strongly affected by the presence of a complex phase in the initial condition, leading to interesting transient effects on the wave packet dynamics.
\section{Self accelerating beams}
\label{sec:3}
\begin{figure*}[t!]
  \includegraphics[width=\linewidth]{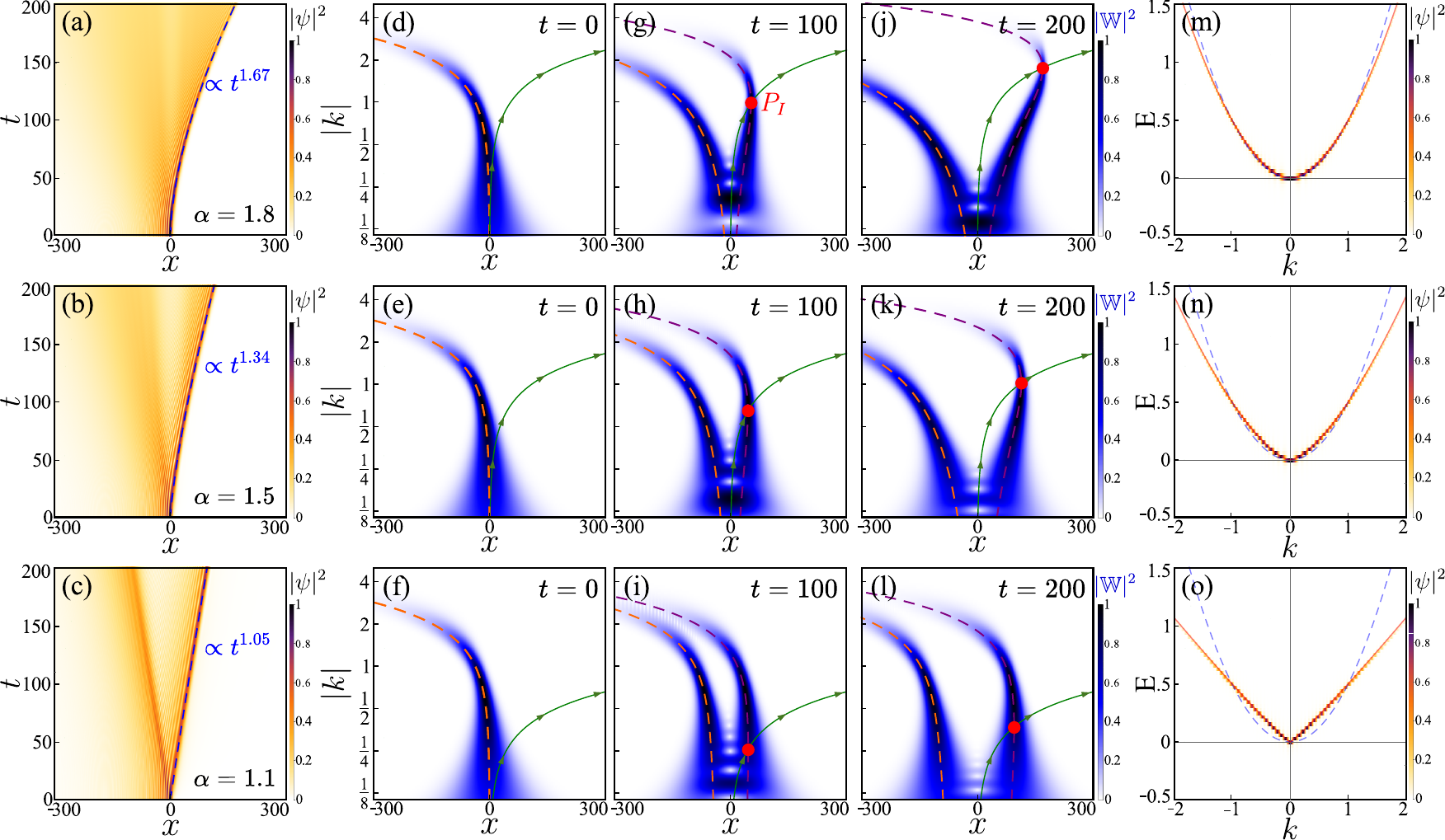}
  \caption{Airy beam dynamics for different fractional orders $\alpha$. Each row corresponds to a different value of $\alpha$. (a)-(c) Wave function density $|\psi_\mathrm{Ai}^{(\alpha)}(x,t)|^2$. The dashed blue line shows the wave packet front trajectory, computed from the extremum mode displacement in Eq.~(\ref{eq:dkextAi}). (d)-(l) Corresponding wavelet energy density $|\mathbb{W}(x,k)|^2$ at selected times. The dashed purple and orange lines are the mode displacement $d_\mathrm{Ai}^{(\alpha)}(k,t)$ computed from Eq.~(\ref{eq:dktAi}). The point $P_I$, around which the self-interference occurs, is shown as a red dot, and the green line stands for its trajectory in the $x$-$k$ phase space. (m)-(o) \textit{Far-field} density $|\psi_\mathrm{Ai}^{(\alpha)}(k,E)|^2$. The field's density follows the dispersion relation $E^{(\alpha)}(k)$, shown as a solid red line. The parabolic dispersion $E^{(\alpha=2)}(k)$ is plotted as a dashed blue line for comparison. Parameters are as follows: $a=0.01$, $b=0.3$, $D_\alpha=0.5$. Supplemental Movie S2 provides an animation of the different Airy beam dynamics with their WT~\cite{noteSM}.
  }
  \label{fig:1}
\end{figure*}
The first self-accelerating beam, the Airy beam, was discovered by Berry and Balazs as a solution of the free one-dimensional Schr\"{o}dinger equation, using an Airy function as initial condition~\cite{berry79a}. However, these beams are not physical solutions as they do not possess a square-integrable wave function, \textit{i.e} they would require an infinite energy to maintain their properties as they propagate. 
Physical approximations of Airy beams have been experimentally realised by including an exponential cut-off on the beam's tail~\cite{siviloglou07a}, in order to ensure its square-integrability. The initial condition thus becomes
\begin{equation}
\psi(x,0)=\mathrm{Ai}(b x)\exp(a x)\,,
\label{eq:Airyinitx}
\end{equation}
where $a$ controls the exponential cut-off of the wave function and $b$ the width of the peaks of the Airy function. 

Other examples of the family of caustic beams that emerge in the catastrophe theory can be constructed from canonical diffraction integrals of codimension $K$~\cite{thom_book72a,berry80a}
\begin{equation}
\xi_K(\boldsymbol{r})=\int_{-\infty}^{+\infty} \textrm{e}^{i V_K(u;\boldsymbol{r})}du \,,
\end{equation}
with the associated potential functions
\begin{equation}
V_K(u;\boldsymbol{r}) = u^{K+2} + \sum_{n=1}^{K} r_n u^n  \,.
\end{equation}
These integrals here depend on one active variable $u$ and on a certain number of control parameters $r_n$. The first example with $K=1$ corresponds to the fold catastrophe, and it is related to the (rescaled) Airy function as 
\begin{equation}
\xi_1(x)=\int_{-\infty}^{+\infty} \textrm{e}^{i (u^3 +u x)}du =\frac{2\pi}{\sqrt[3]{3}}\textrm{Ai}\left(\frac{x}{\sqrt[3]{3}}\right) \,.
\end{equation}
The second example with $K=2$ corresponds to the cusp catastrophe and defines the Pearcey function
\begin{equation}
\xi_2(x,y)=\int_{-\infty}^{+\infty} \textrm{e}^{i (u^4 +u^2 y +u x)}du =\textrm{Pe}(x,y)\,.
\end{equation}
An interesting application of the fold and cusp catastrophes has been recently found as quantum caustics in spin chains systems~\cite{birkby19a}.
The third integral with $K=3$ corresponds to the swallowtail catastrophe
\begin{equation}
\xi_3(x,y,z)=\int_{-\infty}^{+\infty} \textrm{e}^{i (u^5 +u^3 z +u^2 y +u x)}du =\textrm{Sw}(x,y,z)\,.
\end{equation}
One could also construct the four-control parameters integral $\xi_4$, from the Butterfly catastrophe. Thom's theory shows that it only exists seven fundamental types of catastrophe, the three remaining ones, hyperbolic, elliptic and parabolic umbilic catastrophes, are obtained with two active variables~\cite{thom_book72a}.
\\
While multi-dimensional wave packets can be difficult to engineer experimentally, for obvious reasons, one-dimensional versions of high-order catastrophe can be more easily realised. For instance, one-dimensional versions of a finite-energy Pearcey beam and its dual self-accelerating properties have been recently studied~\cite{zangf19a}. The square-integrability of the beam was notably ensured by a symmetrical cut-off on the initial condition, which becomes $\mathrm{Pe}(x,0)\exp(-a x^2)$. One can then similarly construct the next order of one-dimensional accelerating beams, that is a 1D-Swallotail beam $\mathrm{Sw}(x,0,0)$, or any beam of order $n$ arising from the following integral
\begin{equation}
\xi_n(x,\boldsymbol{0})=\int_{-\infty}^{+\infty} \textrm{e}^{i (u^n +u x)}du\,.
\label{eq:intn}
\end{equation}
The integral defined in Eq.~(\ref{eq:intn}) can be notoriously difficult to compute, due to its highly oscillatory integrand. Different methods and algorithms based on sets of differential equations or using contour integration techniques have been developed to evaluate such cupsoid canonical integrals~\cite{connor83a,kirk00a,connor04a}.
\\
The function $\xi_n(x)$ is here defined for an integer-order $n$, but it worth mentioning that such a function could also be defined through a fractional order in its integral representation, similarly to so-called fractional Airy beams~\cite{khonina17a}.

\section{Airy beams in the fractional Schr\"{o}dinger equation}
\label{sec:4}

In this Section, I solve the SFSE for a finite-energy Airy beam as initial condition, see Eq.~(\ref{eq:Airyinitx}), which can be written in momentum space as
\begin{equation}
\psi_0(k)=\mathscr{F}_{k}[\mathrm{Ai}(b x)\exp(a x)]=\frac{\exp\left(\frac{(a - i k)^3}{3 b^3}\right)}{2b \sqrt{2 \pi}}\,.
\label{eq:Airy}
\end{equation}
The solution of Eq.~(\ref{eq:SFSE_k}) is now
\begin{equation}
\psi^{(\alpha)}_\mathrm{Ai}(k,t)=\frac{1}{2b \sqrt{2 \pi}}\exp\left(-i D_\alpha|k|^\alpha t + \frac{(a - i k)^3}{3 b^3}\right)\,.
\label{eq:psiAi}
\end{equation}
The real-space solution $\psi^{(\alpha)}_\mathrm{Ai}(x,t)$ is simply obtained by inverse Fourier transform. Typical dynamics for $|\psi^{(\alpha)}_\mathrm{Ai}(x,t)|^2$ are shown in Fig.~\ref{fig:1}(a)-(c) for three different fractional orders $\alpha$. 
As the fractional order is reduced, the wave front acceleration diminishes, and at long times, the packet undergoes the same reshaping as the one observed in the Gaussian case, see Fig.~\ref{fig:2}.

To understand the wave packet dynamics, I follow the same method as in Section~\ref{sec:2}, first by taking the complex phase of the solution in Eq.~(\ref{eq:psiAi})
\begin{equation}
\phi^{(\alpha)}_\mathrm{Ai}(k,t)=D_\alpha|k|^\alpha t + \frac{3 a^2 k - k^3}{3 b^3}\,,
\end{equation}
from which one obtains the mode displacement
\begin{equation}
d^{(\alpha)}_{\textrm{Ai}} (k,t)=\partial_k \phi^{(\alpha)}_\mathrm{Ai}(k,t) = \alpha D_\alpha t |k|^{\alpha-1}\textrm{sgn}(k) + \frac{a^2 - k^2}{b^3}\,.
\label{eq:dktAi}
\end{equation}
I also compute the WT for the Airy beams with the different fractional orders and show their corresponding wavelet energy densities at three selected times of the evolution in Fig.~\ref{fig:1}(d)-(l). The mode displacement $d^{(\alpha)}_{\textrm{Ai}}$ previously calculated is plotted on top of the wavelet energy density as orange and purple dashed lines, with again an excellent agreement. 

As shown in Ref.~\cite{colas20a}, the presence of fringes in the Airy beam's density $|\psi_\mathrm{Ai}^{(\alpha)}(x,t)|^2$ arises from a transient self-interference of the wave function. The mode displacement $d_\mathrm{Ai}^{(\alpha)}$ is a multivalued function in the $x$-$k$ phase space. A self-interference thus occurs when the wavelet energy density spreads around the point where $d_\mathrm{Ai}^{(\alpha)}$ becomes multivalued. Indeed, for a given position $x$, the wave function can contain two different modes $k$ which overlap in real space and lead to the self-interference. 
The self-interference occurs around a point $P_I$ with coordinates $\{d_\mathrm{Ai}^{(\alpha)}(k_\mathrm{ext}),k_\mathrm{ext} \}$, where $k_\mathrm{ext}$ is the momentum of the extremum mode of the branch, \textit{i.e.} the mode with the largest displacement, or also $\mathrm{max}[d_\mathrm{Ai}^{(\alpha)}]$. The extremum mode is found solving $\partial_k d^{(\alpha)}_{\textrm{Ai}} (k,t)=0$ for $k$, which gives
\begin{equation}
k_\mathrm{ext}=\left(  \frac{b^3}{2} D_\alpha t (\alpha^2-\alpha)\right)^\frac{1}{3-\alpha}\,.
\label{eq:kextAi}
\end{equation}
The displacement of the extremum mode $d^{(\alpha)}_{\textrm{Ai}} (k_\mathrm{ext})$ is found by substituting back Eq.~(\ref{eq:kextAi}) into Eq.~(\ref{eq:dktAi}), which gives
\begin{multline}
d^{(\alpha)}_{\textrm{Ai}} (k_\mathrm{ext},t)= (D_\alpha t)^\frac{2}{3-\alpha}\\
\times\left[ \alpha \left(\frac{b^3}{2}(\alpha^2 -\alpha) \right)^\frac{\alpha-1}{3-\alpha} -\frac{1}{b^3}\left(\frac{b^3}{2}(\alpha^2 -\alpha) \right)^\frac{2}{3-\alpha}\right] +\frac{a^2}{b^3}\,.
\label{eq:dkextAi}
\end{multline}
The point $P_I=\{d_\mathrm{Ai}^{(\alpha)}(k_\mathrm{ext}),k_\mathrm{ext}\}$ is marked as a red dot in the WT panels of Fig.~\ref{fig:1}, and a green line indicates its trajectory in the $x$-$k$ phase space. At long times, the self-interference vanishes as the point $P_I$ drifts away to the region of large momenta, where the wavelet energy density is zero, leaving no signal to participate into the self-interference. This process is faster as $\alpha$ is large, as seen from Fig.~\ref{fig:1}(d)-(l) or from Eq.~(\ref{eq:dkextAi}). Physically, this means that the fringes in the beam's density ``survive'' a longer time as $\alpha$ is reduced to 1. 

Since $d^{(\alpha)}_{\textrm{Ai}} (k_\mathrm{ext})$ measures the distance traveled by the extremum mode, it corresponds to the beam's wave front trajectory in real space. Trajectories for the three cases of Airy beams in Fig.~\ref{fig:1}(a)-(c) and computed from Eq.~(\ref{eq:dkextAi}) are reported as dashed blue lines. For the case of $\alpha=2$ in Eq.~(\ref{eq:dkextAi}), one recovers the usual $t^2$ parabolic acceleration of the Airy beam. 
\\
\\
The last discussion of this Section will concern the splitting of the wave packet at long times, observed either with the Gaussian packet (see Fig.~\ref{fig:2}) or the Airy beam (see Fig.~\ref{fig:1}) and that has been previously discussed~\cite{huangx17a,huangx17b,zangf18a} . It is worth  looking at the effect of lowering the fractional order on the \textit{far-field} density, representing the wave function in both momentum $k$ and energy $E$. The density $|\psi_\mathrm{Ai}^{(\alpha)}(k,E)|^2$ is shown in Fig.~\ref{fig:1}(m)-(o) for the three different Airy cases I have considered so far. In this phase space, the density follows the dispersion relation $E^{(\alpha)}(k)$, here plotted as a red line. One can appreciate its deviation with the parabolic dispersion of the Schr\"{o}dinger equation, shown as a dashed blue line.

The splitting of the beams at long times is only the consequence of the linearisation of the dispersion relation, no matter what the initial condition is. Indeed, even when it contains a complex phase, as for the Airy beam, at long times the effect of the dispersion is dominant on the mode propagation. In Eq.~(\ref{eq:dktAi}), the only time-dependent term is the one that derives from the dispersion, on the LHS. As the dispersion becomes linear ($E^{(\alpha\rightarrow 1)} = |k|$), only two group velocities remain accessible, that are $v_\pm=\pm \alpha D_\alpha$. So if the initial wave function $\psi(k,0)$ is distributed around $k=0$, at long times the packet inevitably splits into two parts, travelling with the same velocity but in opposite directions. In the wavelet picture, this would translate as two vertical lines for the mode displacement on the $x$-$k$ phase space, that is also the signature of non-diffusing wave packets~\cite{colas20a}.

\section{Self accelerating beams in the fractional Schr\"{o}dinger equation}

Let's now generalise the results obtained in the previous Section for the Airy beam, to a general 1D $n\textsuperscript{th}$-order self-accelerating caustic beam, as defined by the integral in Eq.~(\ref{eq:intn}). The method employed in this paper to calculate the mode displacement (see Eqs.~(\ref{eq:dkt}) and~(\ref{eq:dktAi})), which leads to the expression of the acceleration, requires the knowledge of the initial condition in momentum space. Fortunately, the integral $\xi_n(x,\boldsymbol{0})$ possesses a handily Fourier transform. Indeed, one can find that
\begin{equation}
\tilde{\xi}_n(k)=\mathscr{F}_{k}[\xi_n(x,\boldsymbol{0})]=\sqrt{2\pi}\exp(i k^n)\,,
\label{eq:expikn}
\end{equation}
as shown in Appendix~\ref{app:A}. 

It is instructive to first examine the symmetry of the function in Eq.~(\ref{eq:expikn}) in order to deduce some of the beams' properties. When $n$ is even ($n=4,6,8...$), $\tilde{\xi}_n(k)$ possesses both an even real and an even imaginary part, which leads to ${\xi}_n(x)$ having also an even real and an even imaginary part. This follows that the density $|{\xi}_n(x)|^2$ is an even function when $n$ is an even number. Such wave functions, like the Pearcey beam ($n=4$), thus possess a symmetrical double-accelerating wave front~\cite{colas20a}. 
\\
On the other hand, when $n$ is odd ($n=3,5,7...$), $\tilde{\xi}_n(k)$ possesses an even real part and an odd imaginary part, which leads ${\xi}_n(x)$ to be a purely real function with an even real part and an odd real part. The density $|{\xi}_n(x)|^2$ thus has no particular symmetry when $n$ is an odd number. This is why the wave functions of the Airy or the Swallowtail beams ($n=3$ and $5$) only possess a single accelerating wave front.
\\
\\
With $\tilde{\xi}_n(k)$ as initial condition~\cite{note1}, the solution of the SFSE (see Eq.~(\ref{eq:SFSE_k})) now reads
\begin{figure}[t!]
  \includegraphics[width=0.9\linewidth]{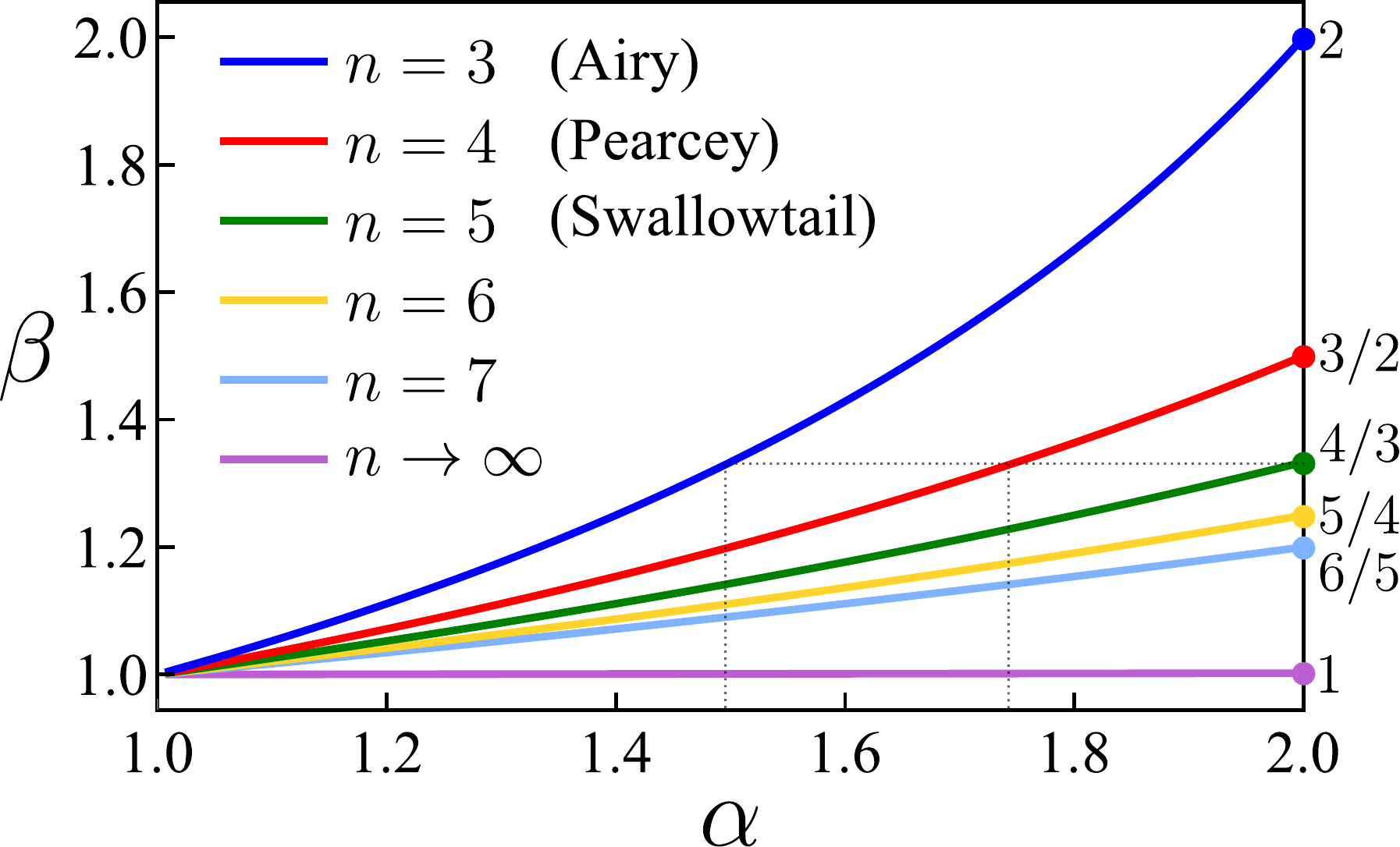}
  \caption{ Exponent of the wave front acceleration $t^\beta$ as function of the fractional order $\alpha$ for self-accelerating beams of different orders $n$. The three cases with $n=3,4$ and $5$ correspond to an Airy, a Pearcey and a Swallotail beam, respectively. These beams can reach the same acceleration for a different fractional order $\alpha$, for example $\beta=4/3$, see the dashed grey line. The dots at $\alpha=2$ indicate the acceleration obtained with the conventional Schr\"{o}dinger equation. The non-accelerating limit $(\beta \rightarrow 1)$ is reached either when $\alpha\rightarrow 1$ or $n\rightarrow \infty$.}
  \label{fig:3}
\end{figure}
\begin{figure}[t!]
  \includegraphics[width=\linewidth]{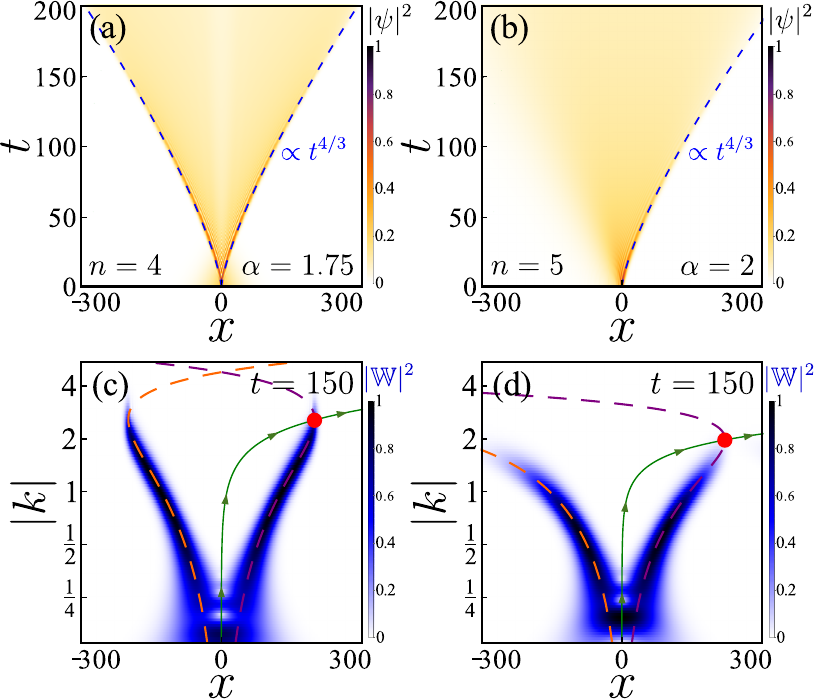}
\caption{ High-order accelerating beams dynamics. The first column corresponds to the case of a Pearcey beam with $n=4$ and $\alpha=1.75$, and the second column to a Swallowtail beam with $n=5$ and $\alpha=2$. (a)-(b) Wave function density $|\psi_n^{(\alpha)}(x,t)|^2$. The dashed blue lines indicate the trajectory of the extremum mode, derived from Eq.~(\ref{eq:dkextn}). (c)-(d) Wavelet energy density $|\mathbb{W}(x,k)|^2$ computed at $t=150$. The dashed orange and purple lines stand for the mode displacement $d^{(\alpha)}_n(k,t)$ derived from Eq.~(\ref{eq:dkn}). The point $P_I$, around which the self-interference occurs, is shown as a red dot, and the green line stands for its trajectory in the $x$-$k$ phase space. Both packets accelerate as $t^{4/3}$. Supplemental Movie S3 provides an animation of the Pearcey and Swallowtail beams dynamics with their WT~\cite{noteSM}.}
  \label{fig:4}
\end{figure}
\begin{equation}
\psi^{(\alpha)}_n(k,t)=\sqrt{N_0}\exp\left(-i \phi^{(\alpha)}_n(k,t) \right)\,,
\end{equation}
with some amplitude $\sqrt{N_0}$ and the associated phase
\begin{equation}
\phi^{(\alpha)}_n(k,t)=D_\alpha|k|^\alpha t - k^n\,.
\end{equation}
The corresponding mode displacement, obtained by differentiating the previous phase with respect to $k$, is
\begin{equation}
d^{(\alpha)}_n(k,t)=\alpha D_\alpha t |k|^{\alpha-1}\textrm{sgn}(k) -n k^{n-1}\,.
\label{eq:dkn}
\end{equation}
The extremum mode is obtained by solving $\partial_k d^{(\alpha)}_n(k,t)=0$ for $k$, giving
\begin{equation}
k_\mathrm{ext}=(n^2-n)^{\frac{1}{\alpha-n}} (D_\alpha t (\alpha^2 -\alpha))^{\frac{1}{n-\alpha}}\,.
\label{eq:kextn}
\end{equation}
Substituting back Eq.~(\ref{eq:kextn}) into Eq.~(\ref{eq:dkn}), one obtains the mode displacement of the extremum mode
\begin{multline}
d^{(\alpha)}_n(k_\mathrm{ext},t)= (D_\alpha t)^\frac{n-1}{n-\alpha}\\
\times\left[ \alpha (n^2 -n)^\frac{\alpha-1}{\alpha-n} (\alpha^2-\alpha)^\frac{\alpha-1}{n-\alpha} \right. \\
\left. - n (n^2 -n)^\frac{n-1}{\alpha-n}  (\alpha^2-\alpha)^\frac{n-1}{n-\alpha}\right]\,,
\label{eq:dkextn}
\end{multline}
which corresponds to the wave front acceleration.

From the previous formula, one finds that the wave packet accelerates as $t^\beta$, with $\beta=(n-1)/(n-\alpha)$. The acceleration parameter $\beta$ as function of the fractional order $\alpha$ is plotted in Fig.~\ref{fig:3} for different beam orders $n$. The values at $\alpha=2$ and marked by a dot correspond to the acceleration obtained for the different beams in the conventional Schr\"{o}dinger equation. One thus retrieves the previously derived $t^{(n-1)/(n-2)}$ acceleration law for general 1D caustic beams~\cite{colas20a}.

From Eq.~(\ref{eq:dkextn}), one finds that the non-accelerating limit ($\beta \rightarrow 1$) is reached either when $\alpha \rightarrow 1$ or when $n \rightarrow \infty$. It also follows from Eq.~(\ref{eq:dkextn}) that it exists different ways to obtain a given acceleration with different types of caustic beams. A given acceleration, for example $\beta=4/3$, can be realised in three ways, (1) with an Airy beam ($n=3$) and a fractional order $\alpha=1.5$, (2) with a Percey beam ($n=4$) and a fractional order $\alpha=1.75$, or (3) with a Swallowtail beam ($n=5$) and a fractional order $\alpha=2$. This Airy beam configuration was already shown in Fig.~\ref{fig:1}(b). Cases of the Pearcey and Swallowtail beams showing the same acceleration are presented in Fig.~\ref{fig:4}. One can note the symmetrical double-accelerating wave front in the Pearcey beam.

The effect of vanishing self-interference is particularly well visible in the supplemental movie S3 that time-animates Fig.~\ref{fig:4}. Indeed, the Pearcey beam possesses here a higher spread in momentum space than the Swallowtail beam, due to their different cut-offs. The self-interference, and thus the fringes in the real space density, are preserved for longer times in the Pearcey beam as the point $P_I$ with coordinates $\{d_n^{(\alpha)}(k_\mathrm{ext}),k_\mathrm{ext} \}$ drifts to large momenta.

\section{Conclusion}
In conclusion, I have studied general one-dimensional self-accelerating caustic beams in the SFSE. I showed that the previously observed effect of the wave packet splitting at long times is a natural consequence of the linearisation of the dispersion relation, which occurs when the fractional order is reduced down to 1. Using a method of analysis based on a combination of spectral techniques, I have calculated the analytical expression for the mode propagation, from which the wave packet acceleration is derived. In both the limits of an infinite caustic order or a fractional order of $1$, the beams loose their accelerating property.
This method of analysis could be employed in the study of other accelerating beams, or beams with a different phase engineering, and with other types of dispersion relations, such as with coupled or periodic systems. 
\appendix
\section{}
\label{app:A}
I want to calculate the Fourier transform of $\xi_n(x,\boldsymbol{0})$, simply written $\xi_n(x)$ in the following, which is defined as
\begin{equation}
\xi_n(x)=\int_{-\infty}^{+\infty} \textrm{e}^{i (u^n +u x)}du\,.
\end{equation}
Using the definition of the Fourier transform, one has
\begin{equation}
\mathscr{F}_{k}[\xi_n(x)]=\int_{-\infty}^{+\infty}\xi_n(x)\textrm{e}^{-i k x}dx\,,
\end{equation}
or also
\begin{equation}
\mathscr{F}_{k}[\xi_n(x)]=\int_{-\infty}^{+\infty}\left(\int_{-\infty}^{+\infty}\textrm{e}^{i (u^n +u x)} du \right)\textrm{e}^{-i k x} dx\,.
\end{equation}
Using Fubini's theorem, one can change the order of integration as
\begin{equation}
\mathscr{F}_{k}[\xi_n(x)]=\int_{-\infty}^{+\infty} \textrm{e}^{i u^n} \left(\int_{-\infty}^{+\infty} \textrm{e}^{i (u -k)x} dx \right) du\,.
\end{equation}
The integral on $dx$ can be seen as a Fourier transform of the identity towards the space $k-u$, which results in a Dirac delta function
\begin{equation}
\mathscr{F}_{k}[\xi_n(x)]=\sqrt{2 \pi}\int_{-\infty}^{+\infty} \textrm{e}^{i u^n} \delta(k-u) du\,,
\end{equation}
which finally leads to
\begin{equation}
\mathscr{F}_{k}[\xi_n(x)]=\sqrt{2 \pi} \textrm{e}^{i k^n}\,.
\end{equation}
\\
\\
\textit{Acknowledgments--}
I thank Fabrice Laussy for critical reading of the manuscript. This research was supported by the Australian Research Council
  Centre of Excellence in Future Low-Energy Electronics Technologies
  (project number CE170100039) and funded by the Australian
  Government.

\begin{thebibliography}{48}%
\makeatletter
\providecommand \@ifxundefined [1]{%
 \@ifx{#1\undefined}
}%
\providecommand \@ifnum [1]{%
 \ifnum #1\expandafter \@firstoftwo
 \else \expandafter \@secondoftwo
 \fi
}%
\providecommand \@ifx [1]{%
 \ifx #1\expandafter \@firstoftwo
 \else \expandafter \@secondoftwo
 \fi
}%
\providecommand \natexlab [1]{#1}%
\providecommand \enquote  [1]{``#1''}%
\providecommand \bibnamefont  [1]{#1}%
\providecommand \bibfnamefont [1]{#1}%
\providecommand \citenamefont [1]{#1}%
\providecommand \href@noop [0]{\@secondoftwo}%
\providecommand \href [0]{\begingroup \@sanitize@url \@href}%
\providecommand \@href[1]{\@@startlink{#1}\@@href}%
\providecommand \@@href[1]{\endgroup#1\@@endlink}%
\providecommand \@sanitize@url [0]{\catcode `\\12\catcode `\$12\catcode
  `\&12\catcode `\#12\catcode `\^12\catcode `\_12\catcode `\%12\relax}%
\providecommand \@@startlink[1]{}%
\providecommand \@@endlink[0]{}%
\providecommand \url  [0]{\begingroup\@sanitize@url \@url }%
\providecommand \@url [1]{\endgroup\@href {#1}{\urlprefix }}%
\providecommand \urlprefix  [0]{URL }%
\providecommand \Eprint [0]{\href }%
\providecommand \doibase [0]{http://dx.doi.org/}%
\providecommand \selectlanguage [0]{\@gobble}%
\providecommand \bibinfo  [0]{\@secondoftwo}%
\providecommand \bibfield  [0]{\@secondoftwo}%
\providecommand \translation [1]{[#1]}%
\providecommand \BibitemOpen [0]{}%
\providecommand \bibitemStop [0]{}%
\providecommand \bibitemNoStop [0]{.\EOS\space}%
\providecommand \EOS [0]{\spacefactor3000\relax}%
\providecommand \BibitemShut  [1]{\csname bibitem#1\endcsname}%
\let\auto@bib@innerbib\@empty
\bibitem [{\citenamefont {Herrmann}(2014)}]{herrmann_book14a}%
  \BibitemOpen
  \bibfield  {author} {\bibinfo {author} {\bibfnamefont {R.}~\bibnamefont
  {Herrmann}},\ }\href@noop {} {\emph {\bibinfo {title} {Fractional calculus:
  an introduction for physicists}}},\ \bibinfo {edition} {2nd}\ ed.\ (\bibinfo
  {publisher} {World Scientific Publishing Company},\ \bibinfo {year}
  {2014})\BibitemShut {NoStop}%
\bibitem [{\citenamefont {Laskin}(2000{\natexlab{a}})}]{laskin00a}%
  \BibitemOpen
  \bibfield  {author} {\bibinfo {author} {\bibfnamefont {N.}~\bibnamefont
  {Laskin}},\ }\href@noop {} {\bibfield  {journal} {\bibinfo  {journal} {Phys.
  Lett. A}\ }\textbf {\bibinfo {volume} {268}},\ \bibinfo {pages} {298}
  (\bibinfo {year} {2000}{\natexlab{a}})}\BibitemShut {NoStop}%
\bibitem [{\citenamefont {Laskin}(2000{\natexlab{b}})}]{laskin00b}%
  \BibitemOpen
  \bibfield  {author} {\bibinfo {author} {\bibfnamefont {N.}~\bibnamefont
  {Laskin}},\ }\href@noop {} {\bibfield  {journal} {\bibinfo  {journal} {Phys.
  Rev. E}\ }\textbf {\bibinfo {volume} {62}},\ \bibinfo {pages} {3135}
  (\bibinfo {year} {2000}{\natexlab{b}})}\BibitemShut {NoStop}%
\bibitem [{\citenamefont {Laskin}(2002)}]{laskin02a}%
  \BibitemOpen
  \bibfield  {author} {\bibinfo {author} {\bibfnamefont {N.}~\bibnamefont
  {Laskin}},\ }\href@noop {} {\bibfield  {journal} {\bibinfo  {journal} {Phys.
  Rev. E}\ }\textbf {\bibinfo {volume} {66}},\ \bibinfo {pages} {056108}
  (\bibinfo {year} {2002})}\BibitemShut {NoStop}%
\bibitem [{\citenamefont {Kilbas}\ \emph {et~al.}(2006)\citenamefont {Kilbas},
  \citenamefont {Srivastava},\ and\ \citenamefont {Trujillo}}]{kilbas_book06a}%
  \BibitemOpen
  \bibfield  {author} {\bibinfo {author} {\bibfnamefont {A.~A.}\ \bibnamefont
  {Kilbas}}, \bibinfo {author} {\bibfnamefont {H.~M.}\ \bibnamefont
  {Srivastava}}, \ and\ \bibinfo {author} {\bibfnamefont {J.~J.}\ \bibnamefont
  {Trujillo}},\ }\href@noop {} {\emph {\bibinfo {title} {Theory and
  Applications of Fractional Differential Equations}}}\ (\bibinfo  {publisher}
  {Elsevier},\ \bibinfo {year} {2006})\BibitemShut {NoStop}%
\bibitem [{\citenamefont {Metzler}\ and\ \citenamefont
  {Klafter}(2000)}]{metzler00a}%
  \BibitemOpen
  \bibfield  {author} {\bibinfo {author} {\bibfnamefont {R.}~\bibnamefont
  {Metzler}}\ and\ \bibinfo {author} {\bibfnamefont {J.}~\bibnamefont
  {Klafter}},\ }\href@noop {} {\bibfield  {journal} {\bibinfo  {journal} {Phys.
  Rep.}\ }\textbf {\bibinfo {volume} {339}},\ \bibinfo {pages} {1} (\bibinfo
  {year} {2000})}\BibitemShut {NoStop}%
\bibitem [{\citenamefont {Sabatier}\ \emph {et~al.}(2007)\citenamefont
  {Sabatier}, \citenamefont {Agrawal},\ and\ \citenamefont
  {Machado}}]{sabatier_book07a}%
  \BibitemOpen
  \bibfield  {author} {\bibinfo {author} {\bibfnamefont {J.}~\bibnamefont
  {Sabatier}}, \bibinfo {author} {\bibfnamefont {O.~P.}\ \bibnamefont
  {Agrawal}}, \ and\ \bibinfo {author} {\bibfnamefont {J.~A.~T.}\ \bibnamefont
  {Machado}},\ }\href@noop {} {\emph {\bibinfo {title} {Advances in Fractional
  Calculus: Theoretical Developments and Applications in Physics and
  Engineering}}}\ (\bibinfo  {publisher} {Springer},\ \bibinfo {year}
  {2007})\BibitemShut {NoStop}%
\bibitem [{\citenamefont {Stickler}(2013)}]{stickler13a}%
  \BibitemOpen
  \bibfield  {author} {\bibinfo {author} {\bibfnamefont {B.~A.}\ \bibnamefont
  {Stickler}},\ }\href@noop {} {\bibfield  {journal} {\bibinfo  {journal}
  {Phys. Rev. E}\ }\textbf {\bibinfo {volume} {88}},\ \bibinfo {pages} {012120}
  (\bibinfo {year} {2013})}\BibitemShut {NoStop}%
\bibitem [{\citenamefont {Longhi}(2015)}]{longhi15a}%
  \BibitemOpen
  \bibfield  {author} {\bibinfo {author} {\bibfnamefont {S.}~\bibnamefont
  {Longhi}},\ }\href@noop {} {\bibfield  {journal} {\bibinfo  {journal} {Opt.
  Lett.}\ }\textbf {\bibinfo {volume} {40}},\ \bibinfo {pages} {1117} (\bibinfo
  {year} {2015})}\BibitemShut {NoStop}%
\bibitem [{\citenamefont {Zhang}\ \emph {et~al.}(2017)\citenamefont {Zhang},
  \citenamefont {Zhang}, \citenamefont {Zhang}, \citenamefont {Ahmed},
  \citenamefont {Zhang}, \citenamefont {Li}, \citenamefont {Belic},\ and\
  \citenamefont {Xiao}}]{zhangd17a}%
  \BibitemOpen
  \bibfield  {author} {\bibinfo {author} {\bibfnamefont {D.}~\bibnamefont
  {Zhang}}, \bibinfo {author} {\bibfnamefont {Y.}~\bibnamefont {Zhang}},
  \bibinfo {author} {\bibfnamefont {Z.}~\bibnamefont {Zhang}}, \bibinfo
  {author} {\bibfnamefont {N.}~\bibnamefont {Ahmed}}, \bibinfo {author}
  {\bibfnamefont {Y.}~\bibnamefont {Zhang}}, \bibinfo {author} {\bibfnamefont
  {F.}~\bibnamefont {Li}}, \bibinfo {author} {\bibfnamefont {M.~R.}\
  \bibnamefont {Belic}}, \ and\ \bibinfo {author} {\bibfnamefont
  {M.}~\bibnamefont {Xiao}},\ }\href@noop {} {\bibfield  {journal} {\bibinfo
  {journal} {Annalen der Physik}\ }\textbf {\bibinfo {volume} {529}},\ \bibinfo
  {pages} {1700149} (\bibinfo {year} {2017})}\BibitemShut {NoStop}%
\bibitem [{\citenamefont {Berry}\ and\ \citenamefont
  {Balazs}(1979)}]{berry79a}%
  \BibitemOpen
  \bibfield  {author} {\bibinfo {author} {\bibfnamefont {M.~V.}\ \bibnamefont
  {Berry}}\ and\ \bibinfo {author} {\bibfnamefont {N.~L.}\ \bibnamefont
  {Balazs}},\ }\href@noop {} {\bibfield  {journal} {\bibinfo  {journal} {Am. J.
  Phys.}\ }\textbf {\bibinfo {volume} {47}},\ \bibinfo {pages} {264} (\bibinfo
  {year} {1979})}\BibitemShut {NoStop}%
\bibitem [{\citenamefont {Siviloglou}\ \emph {et~al.}(2007)\citenamefont
  {Siviloglou}, \citenamefont {Broky}, \citenamefont {Dogariu},\ and\
  \citenamefont {Christodoulides}}]{siviloglou07a}%
  \BibitemOpen
  \bibfield  {author} {\bibinfo {author} {\bibfnamefont {G.~A.}\ \bibnamefont
  {Siviloglou}}, \bibinfo {author} {\bibfnamefont {J.}~\bibnamefont {Broky}},
  \bibinfo {author} {\bibfnamefont {A.}~\bibnamefont {Dogariu}}, \ and\
  \bibinfo {author} {\bibfnamefont {D.~N.}\ \bibnamefont {Christodoulides}},\
  }\href@noop {} {\bibfield  {journal} {\bibinfo  {journal} {Phys. Rev. Lett.}\
  }\textbf {\bibinfo {volume} {99}},\ \bibinfo {pages} {213901} (\bibinfo
  {year} {2007})}\BibitemShut {NoStop}%
\bibitem [{\citenamefont {Siviloglou}\ and\ \citenamefont
  {Christodoulides}(2007)}]{siviloglou07b}%
  \BibitemOpen
  \bibfield  {author} {\bibinfo {author} {\bibfnamefont {G.~A.}\ \bibnamefont
  {Siviloglou}}\ and\ \bibinfo {author} {\bibfnamefont {D.~N.}\ \bibnamefont
  {Christodoulides}},\ }\href@noop {} {\bibfield  {journal} {\bibinfo
  {journal} {Opt. Lett.}\ }\textbf {\bibinfo {volume} {32}},\ \bibinfo {pages}
  {979} (\bibinfo {year} {2007})}\BibitemShut {NoStop}%
\bibitem [{\citenamefont {Ellenbogen}\ \emph {et~al.}(2009)\citenamefont
  {Ellenbogen}, \citenamefont {Voloch-Bloch}, \citenamefont {Ganany-Padowicz},\
  and\ \citenamefont {Arie}}]{ellenbogen09a}%
  \BibitemOpen
  \bibfield  {author} {\bibinfo {author} {\bibfnamefont {T.}~\bibnamefont
  {Ellenbogen}}, \bibinfo {author} {\bibfnamefont {N.}~\bibnamefont
  {Voloch-Bloch}}, \bibinfo {author} {\bibfnamefont {A.}~\bibnamefont
  {Ganany-Padowicz}}, \ and\ \bibinfo {author} {\bibfnamefont {A.}~\bibnamefont
  {Arie}},\ }\href@noop {} {\bibfield  {journal} {\bibinfo  {journal} {Nat.
  Photon.}\ }\textbf {\bibinfo {volume} {3}},\ \bibinfo {pages} {395} (\bibinfo
  {year} {2009})}\BibitemShut {NoStop}%
\bibitem [{\citenamefont {Voloch-Bloch}\ \emph {et~al.}(2013)\citenamefont
  {Voloch-Bloch}, \citenamefont {Lereah}, \citenamefont {Lilach}, \citenamefont
  {Gover},\ and\ \citenamefont {Arie}}]{voloch13a}%
  \BibitemOpen
  \bibfield  {author} {\bibinfo {author} {\bibfnamefont {N.}~\bibnamefont
  {Voloch-Bloch}}, \bibinfo {author} {\bibfnamefont {Y.}~\bibnamefont
  {Lereah}}, \bibinfo {author} {\bibfnamefont {Y.}~\bibnamefont {Lilach}},
  \bibinfo {author} {\bibfnamefont {A.}~\bibnamefont {Gover}}, \ and\ \bibinfo
  {author} {\bibfnamefont {A.}~\bibnamefont {Arie}},\ }\href@noop {} {\bibfield
   {journal} {\bibinfo  {journal} {Nature}\ }\textbf {\bibinfo {volume}
  {494}},\ \bibinfo {pages} {331} (\bibinfo {year} {2013})}\BibitemShut
  {NoStop}%
\bibitem [{\citenamefont {Zhang}\ \emph {et~al.}(2011)\citenamefont {Zhang},
  \citenamefont {Wang}, \citenamefont {Liu}, \citenamefont {Yin}, \citenamefont
  {Lu}, \citenamefont {Chen},\ and\ \citenamefont {Zhang}}]{zhangP11a}%
  \BibitemOpen
  \bibfield  {author} {\bibinfo {author} {\bibfnamefont {P.}~\bibnamefont
  {Zhang}}, \bibinfo {author} {\bibfnamefont {S.}~\bibnamefont {Wang}},
  \bibinfo {author} {\bibfnamefont {Y.}~\bibnamefont {Liu}}, \bibinfo {author}
  {\bibfnamefont {X.}~\bibnamefont {Yin}}, \bibinfo {author} {\bibfnamefont
  {C.}~\bibnamefont {Lu}}, \bibinfo {author} {\bibfnamefont {Z.}~\bibnamefont
  {Chen}}, \ and\ \bibinfo {author} {\bibfnamefont {X.}~\bibnamefont {Zhang}},\
  }\href@noop {} {\bibfield  {journal} {\bibinfo  {journal} {Opt. Lett.}\
  }\textbf {\bibinfo {volume} {36}},\ \bibinfo {pages} {3191} (\bibinfo {year}
  {2011})}\BibitemShut {NoStop}%
\bibitem [{\citenamefont {Baumgartl}\ \emph {et~al.}(2008)\citenamefont
  {Baumgartl}, \citenamefont {Mazilu},\ and\ \citenamefont
  {Dholakia}}]{baumgartl08a}%
  \BibitemOpen
  \bibfield  {author} {\bibinfo {author} {\bibfnamefont {J.}~\bibnamefont
  {Baumgartl}}, \bibinfo {author} {\bibfnamefont {M.}~\bibnamefont {Mazilu}}, \
  and\ \bibinfo {author} {\bibfnamefont {K.}~\bibnamefont {Dholakia}},\
  }\href@noop {} {\bibfield  {journal} {\bibinfo  {journal} {Nat. Photon.}\
  }\textbf {\bibinfo {volume} {2}},\ \bibinfo {pages} {675} (\bibinfo {year}
  {2008})}\BibitemShut {NoStop}%
\bibitem [{\citenamefont {Vettenburg}\ \emph {et~al.}(2014)\citenamefont
  {Vettenburg}, \citenamefont {Dalgarno}, \citenamefont {Nylk}, \citenamefont
  {Coll-Llado}, \citenamefont {Ferrier}, \citenamefont {Cizmar}, \citenamefont
  {Gunn-Moore},\ and\ \citenamefont {Dholakia}}]{vettenburg14a}%
  \BibitemOpen
  \bibfield  {author} {\bibinfo {author} {\bibfnamefont {T.}~\bibnamefont
  {Vettenburg}}, \bibinfo {author} {\bibfnamefont {H.~I.~C.}\ \bibnamefont
  {Dalgarno}}, \bibinfo {author} {\bibfnamefont {J.}~\bibnamefont {Nylk}},
  \bibinfo {author} {\bibfnamefont {C.}~\bibnamefont {Coll-Llado}}, \bibinfo
  {author} {\bibfnamefont {D.~E.~K.}\ \bibnamefont {Ferrier}}, \bibinfo
  {author} {\bibfnamefont {T.}~\bibnamefont {Cizmar}}, \bibinfo {author}
  {\bibfnamefont {F.~J.}\ \bibnamefont {Gunn-Moore}}, \ and\ \bibinfo {author}
  {\bibfnamefont {K.}~\bibnamefont {Dholakia}},\ }\href@noop {} {\bibfield
  {journal} {\bibinfo  {journal} {Nat. Meth.}\ }\textbf {\bibinfo {volume}
  {11}},\ \bibinfo {pages} {541} (\bibinfo {year} {2014})}\BibitemShut
  {NoStop}%
\bibitem [{\citenamefont {Polynkin}\ \emph {et~al.}(2009)\citenamefont
  {Polynkin}, \citenamefont {Kolesik}, \citenamefont {Moloney}, \citenamefont
  {Siviloglou},\ and\ \citenamefont {Christodoulides}}]{polynkin09a}%
  \BibitemOpen
  \bibfield  {author} {\bibinfo {author} {\bibfnamefont {P.}~\bibnamefont
  {Polynkin}}, \bibinfo {author} {\bibfnamefont {M.}~\bibnamefont {Kolesik}},
  \bibinfo {author} {\bibfnamefont {J.~V.}\ \bibnamefont {Moloney}}, \bibinfo
  {author} {\bibfnamefont {G.~A.}\ \bibnamefont {Siviloglou}}, \ and\ \bibinfo
  {author} {\bibfnamefont {D.~N.}\ \bibnamefont {Christodoulides}},\
  }\href@noop {} {\bibfield  {journal} {\bibinfo  {journal} {Science}\ }\textbf
  {\bibinfo {volume} {324}},\ \bibinfo {pages} {229} (\bibinfo {year}
  {2009})}\BibitemShut {NoStop}%
\bibitem [{\citenamefont {Abdollahpour}\ \emph {et~al.}(2010)\citenamefont
  {Abdollahpour}, \citenamefont {Suntsov}, \citenamefont {Papazoglou},\ and\
  \citenamefont {Tzortzakis}}]{abdollahpour10a}%
  \BibitemOpen
  \bibfield  {author} {\bibinfo {author} {\bibfnamefont {D.}~\bibnamefont
  {Abdollahpour}}, \bibinfo {author} {\bibfnamefont {S.}~\bibnamefont
  {Suntsov}}, \bibinfo {author} {\bibfnamefont {D.~G.}\ \bibnamefont
  {Papazoglou}}, \ and\ \bibinfo {author} {\bibfnamefont {S.}~\bibnamefont
  {Tzortzakis}},\ }\href@noop {} {\bibfield  {journal} {\bibinfo  {journal}
  {Phys. Rev. Lett.}\ }\textbf {\bibinfo {volume} {105}},\ \bibinfo {pages}
  {253901} (\bibinfo {year} {2010})}\BibitemShut {NoStop}%
\bibitem [{\citenamefont {Gu}\ and\ \citenamefont {Gbur}(2010)}]{gu10a}%
  \BibitemOpen
  \bibfield  {author} {\bibinfo {author} {\bibfnamefont {Y.}~\bibnamefont
  {Gu}}\ and\ \bibinfo {author} {\bibfnamefont {G.}~\bibnamefont {Gbur}},\
  }\href@noop {} {\bibfield  {journal} {\bibinfo  {journal} {Opt. Lett.}\
  }\textbf {\bibinfo {volume} {35}},\ \bibinfo {pages} {3456} (\bibinfo {year}
  {2010})}\BibitemShut {NoStop}%
\bibitem [{\citenamefont {Nagar}\ and\ \citenamefont
  {Roichman}(2019)}]{nagar19a}%
  \BibitemOpen
  \bibfield  {author} {\bibinfo {author} {\bibfnamefont {H.}~\bibnamefont
  {Nagar}}\ and\ \bibinfo {author} {\bibfnamefont {Y.}~\bibnamefont
  {Roichman}},\ }\href@noop {} {\bibfield  {journal} {\bibinfo  {journal} {Opt.
  Lett.}\ }\textbf {\bibinfo {volume} {44}},\ \bibinfo {pages} {1896} (\bibinfo
  {year} {2019})}\BibitemShut {NoStop}%
\bibitem [{\citenamefont {Thom}(1972)}]{thom_book72a}%
  \BibitemOpen
  \bibfield  {author} {\bibinfo {author} {\bibfnamefont {R.}~\bibnamefont
  {Thom}},\ }\href@noop {} {\emph {\bibinfo {title} {Stabilit\'{e} structurelle
  et morphog\'{e}n\`{e}se : essai d'une th\'{e}orie g\'{e}n\'{e}rale des
  mod\`{e}les.}}}\ (\bibinfo  {publisher} {Reading : W.A. Benjamin},\ \bibinfo
  {year} {1972})\BibitemShut {NoStop}%
\bibitem [{\citenamefont {Berry}\ and\ \citenamefont
  {Upstill}(1980)}]{berry80a}%
  \BibitemOpen
  \bibfield  {author} {\bibinfo {author} {\bibfnamefont {M.~V.}\ \bibnamefont
  {Berry}}\ and\ \bibinfo {author} {\bibfnamefont {C.}~\bibnamefont
  {Upstill}},\ }\href@noop {} {\bibfield  {journal} {\bibinfo  {journal}
  {Progress in Optics}\ }\textbf {\bibinfo {volume} {18}},\ \bibinfo {pages}
  {257} (\bibinfo {year} {1980})}\BibitemShut {NoStop}%
\bibitem [{\citenamefont {Ring}\ \emph {et~al.}(2012)\citenamefont {Ring},
  \citenamefont {Lindberg}, \citenamefont {Mourka}, \citenamefont {Mazilu},
  \citenamefont {Dholakia},\ and\ \citenamefont {Dennis}}]{ring12a}%
  \BibitemOpen
  \bibfield  {author} {\bibinfo {author} {\bibfnamefont {J.~D.}\ \bibnamefont
  {Ring}}, \bibinfo {author} {\bibfnamefont {J.}~\bibnamefont {Lindberg}},
  \bibinfo {author} {\bibfnamefont {A.}~\bibnamefont {Mourka}}, \bibinfo
  {author} {\bibfnamefont {M.}~\bibnamefont {Mazilu}}, \bibinfo {author}
  {\bibfnamefont {K.}~\bibnamefont {Dholakia}}, \ and\ \bibinfo {author}
  {\bibfnamefont {M.~R.}\ \bibnamefont {Dennis}},\ }\href@noop {} {\bibfield
  {journal} {\bibinfo  {journal} {Opt. Express}\ }\textbf {\bibinfo {volume}
  {20}},\ \bibinfo {pages} {18955} (\bibinfo {year} {2012})}\BibitemShut
  {NoStop}%
\bibitem [{\citenamefont {Zannotti}\ \emph {et~al.}(2017)\citenamefont
  {Zannotti}, \citenamefont {Diebel}, \citenamefont {Boguslawski},\ and\
  \citenamefont {Denz}}]{zannotti17a}%
  \BibitemOpen
  \bibfield  {author} {\bibinfo {author} {\bibfnamefont {A.}~\bibnamefont
  {Zannotti}}, \bibinfo {author} {\bibfnamefont {F.}~\bibnamefont {Diebel}},
  \bibinfo {author} {\bibfnamefont {M.}~\bibnamefont {Boguslawski}}, \ and\
  \bibinfo {author} {\bibfnamefont {C.}~\bibnamefont {Denz}},\ }\href@noop {}
  {\bibfield  {journal} {\bibinfo  {journal} {New J. Phys.}\ }\textbf {\bibinfo
  {volume} {19}},\ \bibinfo {pages} {053004} (\bibinfo {year}
  {2017})}\BibitemShut {NoStop}%
\bibitem [{\citenamefont {Zang}\ \emph {et~al.}(2019)\citenamefont {Zang},
  \citenamefont {Wang},\ and\ \citenamefont {Li}}]{zangf19a}%
  \BibitemOpen
  \bibfield  {author} {\bibinfo {author} {\bibfnamefont {F.}~\bibnamefont
  {Zang}}, \bibinfo {author} {\bibfnamefont {Y.}~\bibnamefont {Wang}}, \ and\
  \bibinfo {author} {\bibfnamefont {L.}~\bibnamefont {Li}},\ }\href@noop {} {\
  \textbf {\bibinfo {volume} {15}},\ \bibinfo {pages} {102656} (\bibinfo {year}
  {2019})}\BibitemShut {NoStop}%
\bibitem [{\citenamefont {Huang}\ \emph
  {et~al.}(2017{\natexlab{a}})\citenamefont {Huang}, \citenamefont {Deng},\
  and\ \citenamefont {Fu}}]{huangx17a}%
  \BibitemOpen
  \bibfield  {author} {\bibinfo {author} {\bibfnamefont {X.}~\bibnamefont
  {Huang}}, \bibinfo {author} {\bibfnamefont {Z.}~\bibnamefont {Deng}}, \ and\
  \bibinfo {author} {\bibfnamefont {X.}~\bibnamefont {Fu}},\ }\href@noop {}
  {\bibfield  {journal} {\bibinfo  {journal} {J. Opt. Soc. Am. B}\ }\textbf
  {\bibinfo {volume} {34}},\ \bibinfo {pages} {976} (\bibinfo {year}
  {2017}{\natexlab{a}})}\BibitemShut {NoStop}%
\bibitem [{\citenamefont {Huang}\ \emph
  {et~al.}(2017{\natexlab{b}})\citenamefont {Huang}, \citenamefont {Shi},
  \citenamefont {Dend}, \citenamefont {Bai},\ and\ \citenamefont
  {Fu}}]{huangx17b}%
  \BibitemOpen
  \bibfield  {author} {\bibinfo {author} {\bibfnamefont {X.}~\bibnamefont
  {Huang}}, \bibinfo {author} {\bibfnamefont {X.}~\bibnamefont {Shi}}, \bibinfo
  {author} {\bibfnamefont {Z.}~\bibnamefont {Dend}}, \bibinfo {author}
  {\bibfnamefont {Y.}~\bibnamefont {Bai}}, \ and\ \bibinfo {author}
  {\bibfnamefont {X.}~\bibnamefont {Fu}},\ }\href@noop {} {\bibfield  {journal}
  {\bibinfo  {journal} {Opt. Express}\ }\textbf {\bibinfo {volume} {25}},\
  \bibinfo {pages} {32560} (\bibinfo {year} {2017}{\natexlab{b}})}\BibitemShut
  {NoStop}%
\bibitem [{\citenamefont {Zhang}\ \emph {et~al.}(2019)\citenamefont {Zhang},
  \citenamefont {Zhang}, \citenamefont {Wu}, \citenamefont {Li}, \citenamefont
  {Pierangeli}, \citenamefont {Gao},\ and\ \citenamefont {Fan}}]{zhangl19a}%
  \BibitemOpen
  \bibfield  {author} {\bibinfo {author} {\bibfnamefont {L.}~\bibnamefont
  {Zhang}}, \bibinfo {author} {\bibfnamefont {X.}~\bibnamefont {Zhang}},
  \bibinfo {author} {\bibfnamefont {H.}~\bibnamefont {Wu}}, \bibinfo {author}
  {\bibfnamefont {C.}~\bibnamefont {Li}}, \bibinfo {author} {\bibfnamefont
  {D.}~\bibnamefont {Pierangeli}}, \bibinfo {author} {\bibfnamefont
  {Y.}~\bibnamefont {Gao}}, \ and\ \bibinfo {author} {\bibfnamefont
  {D.}~\bibnamefont {Fan}},\ }\href@noop {} {\bibfield  {journal} {\bibinfo
  {journal} {Opt. Express}\ }\textbf {\bibinfo {volume} {27}},\ \bibinfo
  {pages} {027936} (\bibinfo {year} {2019})}\BibitemShut {NoStop}%
\bibitem [{\citenamefont {Colas}\ \emph {et~al.}(2020)\citenamefont {Colas},
  \citenamefont {Laussy},\ and\ \citenamefont {Davis}}]{colas20a}%
  \BibitemOpen
  \bibfield  {author} {\bibinfo {author} {\bibfnamefont {D.}~\bibnamefont
  {Colas}}, \bibinfo {author} {\bibfnamefont {F.}~\bibnamefont {Laussy}}, \
  and\ \bibinfo {author} {\bibfnamefont {M.~J.}\ \bibnamefont {Davis}},\
  }\href@noop {} {\bibfield  {journal} {\bibinfo  {journal} {Phys. Rev. Res.}\
  }\textbf {\bibinfo {volume} {2}},\ \bibinfo {pages} {023337} (\bibinfo {year}
  {2020})}\BibitemShut {NoStop}%
\bibitem [{\citenamefont {Colas}\ and\ \citenamefont
  {Laussy}(2016)}]{colas16a}%
  \BibitemOpen
  \bibfield  {author} {\bibinfo {author} {\bibfnamefont {D.}~\bibnamefont
  {Colas}}\ and\ \bibinfo {author} {\bibfnamefont {F.~P.}\ \bibnamefont
  {Laussy}},\ }\href@noop {} {\bibfield  {journal} {\bibinfo  {journal} {Phys.
  Rev. Lett.}\ }\textbf {\bibinfo {volume} {116}},\ \bibinfo {pages} {026401}
  (\bibinfo {year} {2016})}\BibitemShut {NoStop}%
\bibitem [{\citenamefont {Colas}\ \emph {et~al.}(2018)\citenamefont {Colas},
  \citenamefont {Laussy},\ and\ \citenamefont {Davis}}]{colas18a}%
  \BibitemOpen
  \bibfield  {author} {\bibinfo {author} {\bibfnamefont {D.}~\bibnamefont
  {Colas}}, \bibinfo {author} {\bibfnamefont {F.}~\bibnamefont {Laussy}}, \
  and\ \bibinfo {author} {\bibfnamefont {M.~J.}\ \bibnamefont {Davis}},\
  }\href@noop {} {\bibfield  {journal} {\bibinfo  {journal} {Phys. Rev. Lett.}\
  }\textbf {\bibinfo {volume} {121}},\ \bibinfo {pages} {055302} (\bibinfo
  {year} {2018})}\BibitemShut {NoStop}%
\bibitem [{\citenamefont {Colas}\ \emph {et~al.}(2019)\citenamefont {Colas},
  \citenamefont {Laussy},\ and\ \citenamefont {Davis}}]{colas19a}%
  \BibitemOpen
  \bibfield  {author} {\bibinfo {author} {\bibfnamefont {D.}~\bibnamefont
  {Colas}}, \bibinfo {author} {\bibfnamefont {F.}~\bibnamefont {Laussy}}, \
  and\ \bibinfo {author} {\bibfnamefont {M.~J.}\ \bibnamefont {Davis}},\
  }\href@noop {} {\bibfield  {journal} {\bibinfo  {journal} {Phys. Rev. B}\
  }\textbf {\bibinfo {volume} {99}},\ \bibinfo {pages} {214301} (\bibinfo
  {year} {2019})}\BibitemShut {NoStop}%
\bibitem [{\citenamefont {\c{S}. Bayin}(2016)}]{bayin16a}%
  \BibitemOpen
  \bibfield  {author} {\bibinfo {author} {\bibfnamefont {S.}~\bibnamefont
  {\c{S}. Bayin}},\ }\href@noop {} {\bibfield  {journal} {\bibinfo  {journal}
  {J. Math. Phys.}\ }\textbf {\bibinfo {volume} {57}},\ \bibinfo {pages}
  {123501} (\bibinfo {year} {2016})}\BibitemShut {NoStop}%
\bibitem [{not({\natexlab{a}})}]{note2}%
  \BibitemOpen
  \href@noop {} {} ({\natexlab{a}}),\ \bibinfo {note} {the superscript on
  $X^{(\alpha)}$ indicates the fractional order for a quantity
  $X$.}\BibitemShut {Stop}%
\bibitem [{\citenamefont {Neto}\ \emph {et~al.}(2009)\citenamefont {Neto},
  \citenamefont {Guinea}, \citenamefont {Peres}, \citenamefont {Novoselov},\
  and\ \citenamefont {Geim}}]{castronetoa09}%
  \BibitemOpen
  \bibfield  {author} {\bibinfo {author} {\bibfnamefont {A.~H.~C.}\
  \bibnamefont {Neto}}, \bibinfo {author} {\bibfnamefont {F.}~\bibnamefont
  {Guinea}}, \bibinfo {author} {\bibfnamefont {N.~M.~R.}\ \bibnamefont
  {Peres}}, \bibinfo {author} {\bibfnamefont {K.~S.}\ \bibnamefont
  {Novoselov}}, \ and\ \bibinfo {author} {\bibfnamefont {A.~K.}\ \bibnamefont
  {Geim}},\ }\href@noop {} {\bibfield  {journal} {\bibinfo  {journal} {Rev.
  Mod. Phys.}\ }\textbf {\bibinfo {volume} {81}},\ \bibinfo {pages} {109}
  (\bibinfo {year} {2009})}\BibitemShut {NoStop}%
\bibitem [{\citenamefont {Utsunomiya}\ \emph {et~al.}(2008)\citenamefont
  {Utsunomiya}, \citenamefont {Tian}, \citenamefont {Roumpos}, \citenamefont
  {Lai}, \citenamefont {Kumada}, \citenamefont {Fujisawa}, \citenamefont
  {Kuwata-Gonokami}, \citenamefont {L\"offler}, \citenamefont {H\"ofling},
  \citenamefont {Forchel},\ and\ \citenamefont {Yamamoto}}]{utsunomiya08a}%
  \BibitemOpen
  \bibfield  {author} {\bibinfo {author} {\bibfnamefont {S.}~\bibnamefont
  {Utsunomiya}}, \bibinfo {author} {\bibfnamefont {L.}~\bibnamefont {Tian}},
  \bibinfo {author} {\bibfnamefont {G.}~\bibnamefont {Roumpos}}, \bibinfo
  {author} {\bibfnamefont {C.~W.}\ \bibnamefont {Lai}}, \bibinfo {author}
  {\bibfnamefont {N.}~\bibnamefont {Kumada}}, \bibinfo {author} {\bibfnamefont
  {T.}~\bibnamefont {Fujisawa}}, \bibinfo {author} {\bibfnamefont
  {M.}~\bibnamefont {Kuwata-Gonokami}}, \bibinfo {author} {\bibfnamefont
  {A.}~\bibnamefont {L\"offler}}, \bibinfo {author} {\bibfnamefont
  {S.}~\bibnamefont {H\"ofling}}, \bibinfo {author} {\bibfnamefont
  {A.}~\bibnamefont {Forchel}}, \ and\ \bibinfo {author} {\bibfnamefont
  {Y.}~\bibnamefont {Yamamoto}},\ }\href@noop {} {\bibfield  {journal}
  {\bibinfo  {journal} {Nat. Phys.}\ }\textbf {\bibinfo {volume} {4}},\
  \bibinfo {pages} {700} (\bibinfo {year} {2008})}\BibitemShut {NoStop}%
\bibitem [{not({\natexlab{b}})}]{noteSM}%
  \BibitemOpen
  \href@noop {} {} ({\natexlab{b}}),\ \bibinfo {note} {see Supplemental at [URL
  will be inserted by publisher] for three videos, consisting in time-animated
  version of Figures 1, 2 and 4.}\BibitemShut {Stop}%
\bibitem [{\citenamefont {Sonin}(2016)}]{sonin_book16a}%
  \BibitemOpen
  \bibfield  {author} {\bibinfo {author} {\bibfnamefont {E.~B.}\ \bibnamefont
  {Sonin}},\ }\href@noop {} {\emph {\bibinfo {title} {Dynamics of Quantised
  Vortices in Superfluids}}}\ (\bibinfo  {publisher} {Cambridge University
  Press},\ \bibinfo {year} {2016})\BibitemShut {NoStop}%
\bibitem [{\citenamefont {Debnath}\ and\ \citenamefont
  {Shah}(2015)}]{debnath_book15a}%
  \BibitemOpen
  \bibfield  {author} {\bibinfo {author} {\bibfnamefont {L.}~\bibnamefont
  {Debnath}}\ and\ \bibinfo {author} {\bibfnamefont {F.~A.}\ \bibnamefont
  {Shah}},\ }\href@noop {} {\emph {\bibinfo {title} {Wavelet Transforms and
  Their Applications}}},\ \bibinfo {edition} {2nd}\ ed.\ (\bibinfo  {publisher}
  {Birkh{\u a}user},\ \bibinfo {year} {2015})\BibitemShut {NoStop}%
\bibitem [{\citenamefont {Kirkby}\ \emph {et~al.}(2019)\citenamefont {Kirkby},
  \citenamefont {Mumford},\ and\ \citenamefont {O'Dell}}]{birkby19a}%
  \BibitemOpen
  \bibfield  {author} {\bibinfo {author} {\bibfnamefont {W.}~\bibnamefont
  {Kirkby}}, \bibinfo {author} {\bibfnamefont {J.}~\bibnamefont {Mumford}}, \
  and\ \bibinfo {author} {\bibfnamefont {D.~H.~J.}\ \bibnamefont {O'Dell}},\
  }\href@noop {} {\bibfield  {journal} {\bibinfo  {journal} {Phys. Rev. Res.}\
  }\textbf {\bibinfo {volume} {1}},\ \bibinfo {pages} {033135} (\bibinfo {year}
  {2019})}\BibitemShut {NoStop}%
\bibitem [{\citenamefont {Connor}\ \emph {et~al.}(1983)\citenamefont {Connor},
  \citenamefont {Curtis},\ and\ \citenamefont {Farrelly}}]{connor83a}%
  \BibitemOpen
  \bibfield  {author} {\bibinfo {author} {\bibfnamefont {J.~N.~L.}\
  \bibnamefont {Connor}}, \bibinfo {author} {\bibfnamefont {P.~R.}\
  \bibnamefont {Curtis}}, \ and\ \bibinfo {author} {\bibfnamefont
  {D.}~\bibnamefont {Farrelly}},\ }\href@noop {} {\bibfield  {journal}
  {\bibinfo  {journal} {Molecular Physics}\ }\textbf {\bibinfo {volume} {48}},\
  \bibinfo {pages} {1305} (\bibinfo {year} {1983})}\BibitemShut {NoStop}%
\bibitem [{\citenamefont {Kirk}\ \emph {et~al.}(2000)\citenamefont {Kirk},
  \citenamefont {Connor},\ and\ \citenamefont {Hobbs}}]{kirk00a}%
  \BibitemOpen
  \bibfield  {author} {\bibinfo {author} {\bibfnamefont {N.~P.}\ \bibnamefont
  {Kirk}}, \bibinfo {author} {\bibfnamefont {J.~N.~L.}\ \bibnamefont {Connor}},
  \ and\ \bibinfo {author} {\bibfnamefont {C.~A.}\ \bibnamefont {Hobbs}},\
  }\href@noop {} {\bibfield  {journal} {\bibinfo  {journal} {Comp. Phys.
  Comm.}\ }\textbf {\bibinfo {volume} {132}},\ \bibinfo {pages} {142} (\bibinfo
  {year} {2000})}\BibitemShut {NoStop}%
\bibitem [{\citenamefont {Connor}\ and\ \citenamefont
  {Hobbs}(2004)}]{connor04a}%
  \BibitemOpen
  \bibfield  {author} {\bibinfo {author} {\bibfnamefont {J.~N.~L.}\
  \bibnamefont {Connor}}\ and\ \bibinfo {author} {\bibfnamefont {C.~A.}\
  \bibnamefont {Hobbs}},\ }\href@noop {} {\bibfield  {journal} {\bibinfo
  {journal} {Khimicheskaya Fizika}\ }\textbf {\bibinfo {volume}
  {23}},\ \bibinfo {pages} {13} (\bibinfo {year} {2004})}\BibitemShut {NoStop}%
\bibitem [{\citenamefont {Khonina}\ and\ \citenamefont
  {Ustinov}(2017)}]{khonina17a}%
  \BibitemOpen
  \bibfield  {author} {\bibinfo {author} {\bibfnamefont {S.~N.}\ \bibnamefont
  {Khonina}}\ and\ \bibinfo {author} {\bibfnamefont {A.~V.}\ \bibnamefont
  {Ustinov}},\ }\href@noop {} {\bibfield  {journal} {\bibinfo  {journal} {J.
  Opt. Soc. Am. A}\ }\textbf {\bibinfo {volume} {34}},\ \bibinfo {pages} {1991}
  (\bibinfo {year} {2017})}\BibitemShut {NoStop}%
\bibitem [{\citenamefont {Zang}\ \emph {et~al.}(2018)\citenamefont {Zang},
  \citenamefont {Wang},\ and\ \citenamefont {Li}}]{zangf18a}%
  \BibitemOpen
  \bibfield  {author} {\bibinfo {author} {\bibfnamefont {F.}~\bibnamefont
  {Zang}}, \bibinfo {author} {\bibfnamefont {Y.}~\bibnamefont {Wang}}, \ and\
  \bibinfo {author} {\bibfnamefont {L.}~\bibnamefont {Li}},\ }\href@noop {}
  {\bibfield  {journal} {\bibinfo  {journal} {Opt. Express}\ }\textbf {\bibinfo
  {volume} {26}},\ \bibinfo {pages} {023740} (\bibinfo {year}
  {2018})}\BibitemShut {NoStop}%
\bibitem [{not({\natexlab{c}})}]{note1}%
  \BibitemOpen
  \href@noop {} {} ({\natexlab{c}}),\ \bibinfo {note} {to simplify
  calculations, the truncation on the beam's density is not included here. The
  obtained mode displacement thus corresponds to infinite-energy beams.
  However, it is a good approximation for the finite-energy cases since the
  cut-off values are typically small, such as in Eq.~\ref{eq:dkextAi} where $a
  \ll 1$.}\BibitemShut {Stop}%
\end{thebibliography}
\end{document}